\documentclass[12pt,letterpaper]{article}
\usepackage{jheppub}
\usepackage{verbatim}
\usepackage{slashed}

\usepackage{graphicx}
\usepackage{subfigure}
\input{epsf}
\usepackage{epsfig}
\usepackage{epstopdf}
\usepackage{amsthm}
\usepackage{tikz}
\usepackage{soul}

\def\({\left(} \def\){\right)}
\def\[{\left[} \def\]{\right]}

\newcommand{\be}{\begin{equation}}
\newcommand{\ee}{\end{equation}}
\newcommand{\bea}{\begin{eqnarray}}
\newcommand{\eea}{\end{eqnarray}}
\newcommand{\ba}{\begin{eqnarray}}
\newcommand{\ea}{\end{eqnarray}}

\newcommand{\beq}{\begin{equation}}
\newcommand{\eeq}{\end{equation}}
\newcommand{\beqa}{\begin{eqnarray}}
\newcommand{\eeqa}{\end{eqnarray}}
\newcommand{\beqar}{\begin{eqnarray*}}
\newcommand{\eeqar}{\end{eqnarray*}}

\newcommand{\eg}{{\it e.g.}\ }
\newcommand{\ie}{{\it i.e.}\ }
\title{On string theory on (deformed)  $AdS_3\times \mathbb{T}^3$}

\author[a]{Soumangsu Chakraborty} 
\author[b]{and Amit Giveon}

\affiliation[a]{Department of Physics,
Center for Cosmology and AstroParticle Physics (CCAPP)\\
The Ohio State University,
191 W Woodruff Ave, Columbus, OH 43210, USA}

\affiliation[b]{Racah Institute of Physics, The Hebrew University, Jerusalem, 91904, Israel}

\emailAdd{soumangsuchakraborty@gmail.com}
\emailAdd{giveon@mail.huji.ac.il}

\abstract{We revisit the fermionic string theory on $AdS_3\times\mathcal{N}$ with $k=1$, and its single-trace $T\bar T$ deformation, with a focus on the  $(2,2)$ superstring on (deformed) $AdS_3\times \mathbb{T}^3$. 
In a certain limit, it is dual to the symmetric product of the ($T\bar T$-deformed) SCFT$_2$ on $\mathbb{R}\times \mathbb{T}^3$.
We present the winding-one delta-function normalizable worldsheet operators which, in the $k=1$ decoupling limit, correspond to those of $\mathbb{R}\times \mathbb{T}^3$ in spacetime. We then demonstrate how their properties in string theory reproduce those of $\mathbb{R}\times\mathbb{T}^3$, or more generally, of a ($T\bar{T}$-deformed) $\mathbb{R}\times\mathcal{N}$ seed of the boundary theory.
}

\begin{document}
\maketitle

\section{Introduction}\label{intro}

In \cite{Balthazar:2021xeh}, it was pointed out that superstring theory on $AdS_3 \times \mathcal{N}$ with (NS,NS) $B$-field background and $R_{AdS}/l_s=\sqrt k<1$ is dual to a symmetric product CFT$_2$ deformed by an operator in the $\mathbb{Z}_2$ twisted sector.

Recently, \cite{Chakraborty:2025nlb}, the analysis of \cite{Balthazar:2021xeh} was generalized to $k>1$. There, it was shown that the resulting picture matches the one discussed in the bosonic case in \cite{Eberhardt:2021vsx}.
The proposal of~\cite{Eberhardt:2021vsx} was that the physics at energy scales that remain finite in the limit where the spacetime central charge goes to infinity could be described by a symmetric product with a marginal perturbation in the $\mathbb{Z}_2$ twisted sector, like for $k<1$.
It was also argued in \cite{Chakraborty:2025nlb} that in the continuation to the critical case, $k\to 1$ \cite{Giveon:2005mi}, the $\mathbb{Z}_2$ twisted deformation remains non-trivial, and the perturbative string spectrum contains both discrete states (short-strings) and the continuum (long-strings).

More recently, it was argued in \cite{Eberhardt:2025sbi} that for $k=1$ ($k=3$ in the bosonic string), one can define a new string theory on $AdS_3 \times \mathcal{N}$, that only includes the (spectral flowed) continuous representations, a.k.a delta-function normalizable operators with non-zero winding, and is dual to a symmetric orbifold CFT$_2$ with an $\mathbb{R}\times\mathcal{N}$ seed.~\footnote{See also \cite{Giveon:2005mi,Giribet:2018ada,Gaberdiel:2018rqv,Gaberdiel:2024dva,Giribet:2020mkc}, for previous attempts in string theory on $AdS_3$ with $k=1$.}

In this paper, we describe the results of \cite{Eberhardt:2025sbi} from the perspective of \cite{Balthazar:2021xeh,Eberhardt:2021vsx,Chakraborty:2025nlb}, and extend them to a symmetric product of $T\bar T$-deformed CFT$_2$. 
On the way, we answer a couple of questions that we describe next.

The first question is the following.
The seed of the long-strings symmetric product, $\mathbb{R}\times \mathcal{N}$, has a free scalar field, $\phi$, parametrizing $\mathbb{R}$, that gives rise to a holomorphic current, $\partial_x \phi$. 
But in the standard string theory on $AdS_3\times{\cal N}$, $\partial_{\bar x}\partial_x\phi\neq 0$, due to the presence of the $\mathbb{Z}_2$ twisted wall,~\cite{Balthazar:2021xeh,Chakraborty:2025nlb}.
This raises the question: {\it how does one obtain such a holomorphic operator in a holographic dual without a wall, namely, in $\mathbb{R}\times \mathcal{N}$, from its construction in string theory?} 
A related question concerns the enhancement of other symmetries, such as supersymmetry. 
Namely, how does the $(4,4)$ supersymmetry of $(\mathbb{R}\times \mathbb{T}^3)^p/S_p$  arises from the superstring on $AdS_3\times \mathbb{T}^3$?

In this paper, we present the limiting procedure (which is possible only when $k=1$), that gives rise to the on-shell delta-function normalizable worldsheet operator corresponding to $\partial_x \phi$ in the winding-one sector, and show that 
$\partial_{\bar x}\partial_x\phi=0$ in the $k=1$ decoupled limit. Moreover, in the particular case where $\mathcal{N}$ allows an $N=(2,2)$ supersymmetry in spacetime, we present the worldsheet construction of the operators corresponding to the fermionic superpartners of $\partial_x \phi$, in the seed of the long-strings decoupled theory,~$\mathbb{R}\times{\cal N}$.

The second question is the following.
According to the standard holographic dictionary, the normalizable and delta-function normalizable states on the string theory side correspond to physical states in the Hilbert space of the boundary CFT$_2$, and the non-normalizable string theory operators correspond to local operators of the boundary theory. The new string theory on $AdS_3$ with $k=1$ ($k=3$ for bosonic strings), \cite{Eberhardt:2025sbi}, includes the delta-function normalizable states with non-zero winding, regards their positive and negative (real) radial momenta as being independent, and discards both the normalizable and non-normalizable modes. This raises an obvious question: {\it how to account for the local operators of the boundary field theory?} 

In this paper, we show that non-normalizable operators of the standard string theory on $AdS_3\times \mathcal{N}$ have in the $k=1$  limit local representatives in the winding-one sector of the delta-function normalizable branch, which account for the local operators of the boundary theory in the limit. We explicitly construct the winding-one representatives of conserved currents, the stress-energy tensor, and a couple of single-trace irrelevant operators ($T\bar{T}$ and $J\bar{T}$), and argue that they have the desired properties.
The $AdS_3$ construction at $k=1$ can readily be generalized to its single-trace $T\bar{T}$ deformation, and its generalizations.

Many of our manipulations and results apply to generic fermionic string theory on (deformed) $AdS_3\times{\cal N}$ with $k=1$, and our presentation will thus be general, when appropriate. Nevertheless,
for concreteness, in section \ref{sec2} (section \ref{sec3}) we shall sometimes focus on the superstring on (deformed) $AdS_3\times \mathbb{T}^3$   (${\cal M}_3\times \mathbb{T}^3$),
and conclude with a more general discussion in section \ref{diss}. In the appendices, we present some technical details and facts.

This note is an extension of the note added section in \cite{Chakraborty:2025nlb}.
Below, we use extensively the results and notation of  \cite{Chakraborty:2025nlb,Balthazar:2021xeh}, which include many additional details and references to earlier literature. Thus, this note is best read in conjunction with those papers.

\section{String theory on $AdS_3\times \mathcal{N}$ and on $AdS_3\times \mathbb{T}^3$}\label{sec2}

In this section, after recalling some properties of the fermionic string theory on $AdS_3\times \mathcal{N}$ (in subsection \ref{sec2.1}),
we restrict to the case $k=1$ (in subsection \ref{sec2.2}). Then, we describe the properties of three special operators (in subsections \ref{sec2.3},~\ref{sec2.4} and~\ref{sec2.5}, respectively), and mention more examples in the superstring on $AdS_3\times \mathbb{T}^3$ (in subsection \ref{sec2.6}).  
Finally, we present the single-trace $T\bar T$ and $J\bar T$ operators for the fermionic string on $AdS_3\times{\cal N}$ in the $k=1$ limit (in subsection \ref{sec2.7}).

\subsection{Fermionic string theory on $AdS_3 \times \mathcal{N}$}\label{sec2.1}

To describe the fermionic string on $AdS_3$ with $k$ units of (NS,NS) $B$-flux,
we need~\cite{Giveon:1998ns} to choose a unitary compact SCFT $\cal N$, whose central charge is determined by the criticality of the full background,
\be\label{k}
\left(3+{6\over k}\right)+{3\over 2}+c_{\cal N}=15~,
\ee
where the first two terms on the l.h.s. are the central charges of the bosonic $SL(2,\mathbb{R})$ WZW model at level $k+2$, and three fermions $\psi^a$, $a=1,2,3$, that transform in the adjoint representation of level $-2$ $SL(2,\mathbb{R})$ current algebra, and $c_{\cal N}$ is the central charge of the SCFT ${\cal N}$.

The theory on a single long-string winding once on the spatial circle and propagating in the bulk of $AdS_3$ is described \cite{Seiberg:1999xz} by the SCFT
\be\label{sw}
M_{6k}=\mathbb{R}_\phi\times{\cal N}~,
\ee
with central charge $6k$, where $\mathbb{R}_\phi$ describes the radial motion of the string, and is given by a linear dilaton CFT for
a field $\phi$. It also includes a free fermion $\psi_\phi$, which is related to $\phi$ by an emergent supersymmetry of the long-string theory, \eqref{sw}.
The linear dilaton slope is
(as can be checked using \eqref{k})
\be\label{ql}
Q_\ell=\sqrt{2\over k}(1-k)~,
\ee
such that the string coupling for the long-strings is
\be\label{glong}
g_\ell(\phi)\sim\exp\left(-{1\over 2}Q_\ell\phi\right)~,
\ee
where $\phi\to\infty$ amounts to the boundary of $AdS_3$.~\footnote{To see that the minus sign in \eqref{glong} is correct for the  $Q_\ell$ in \eqref{ql}, see \eg appendix A of~\cite{Giveon:2005mi}.}

Multiple long-strings propagating in the bulk are (approximately) free, and are described by the symmetric product SCFT, whose seed is $M_{6k}$.
In perturbative superstring theory on $AdS_3\times{\cal N}$ with parametrically small string coupling $g_s$, which amounts to the physics near $p$ fundamental strings, the long-strings spectrum indeed has the pattern of
\be\label{symmprod}
(M_{6k})^p/S_p~,
\ee
where $p\sim 1/g_s^2$, ~\cite{Argurio:2000tb,Giveon:2005mi}.

The theory \eqref{symmprod} is singular, due to the presence of the linear dilaton, and is blind to the existence of discrete, normalizable states, as well as to non-perturbative black hole states (whose energies go like $1/g_s^2$).
In~\cite{Chakraborty:2025nlb}, following~\cite{Balthazar:2021xeh} and~\cite{Eberhardt:2021vsx}, it was argued that this symmetric product structure is modified by 
a $\mathbb{Z}_2$ twisted marginal deformation in the full theory. For $k<1$, the case discussed in~\cite{Balthazar:2021xeh}, $g_\ell$ is small at large $\phi$, \eqref{glong}, and this description can be extended to arbitrarily high energies; consequently, it provides the exact holographic dual of \eg the superstring on $AdS_3\times \mathbb{S}^1\times SU(2)_n/U(1)$ (which has $k={n\over n+1}<1$). On the other hand, for $k>1$, it only captures the perturbative string states (since the long-strings coupling $g_\ell$ is large at large $\phi$,~\eqref{glong}).

\subsection{Fermionic string theory on $AdS_3\times{\cal N}$ with $k=1$}\label{sec2.2}

A natural question that one may raise is what happens at the transition point between the $k<1$ and $k>1$ regimes, namely, for $k=1$.
This can be discussed from the perspective of the boundary CFT, or from that of the bulk string theory on $AdS_3$.
In the following, we describe the two perspectives on this question, starting with the former in this subsection.

A useful starting point is the theory with $k<1$ studied in \cite{Balthazar:2021xeh}, which, as mentioned above, is better defined mathematically than that with $k>1$. The dual CFT is in this case a symmetric product with seed $\mathbb{R}_\phi\times\mathcal{N}$ with a $\mathbb{Z}_2$ twisted marginal wall that prevents the system from exploring the strong coupling region. One may expect that by focusing on physics far from the wall, \ie by restricting to large $\phi$ in \eqref{sw}, we get a theory in which the wall is absent.
This can be done by formally setting the coefficient of the $\mathbb{Z}_2$ twisted deformation, $\mu$, to zero.

For generic $k$, this attempt fails, due to the non-vanishing linear dilaton slope $Q_\ell$ given in eq. \eqref{ql}. This linear dilaton causes amplitudes of generic (real) momentum modes to be ill-defined. If one starts with the theory with the $\mathbb{Z}_2$ twisted wall present, and tries to remove it by sending its location to $\phi\to -\infty$,
the limit is singular.\footnote{One can define bulk amplitudes, in the sense of \cite{Aharony:2004xn}, but those involve in general, operators with imaginary momentum. Moreover, the sets of bulk amplitudes (that are characterized by the sum rules satisfied by the momenta) are different at different genera (orders of string perturbation theory), and have contributions due to LSZ poles \cite{Aharony:2004xn}.}
This is related to the fact that for general $k$ the theory is not translationally invariant in $\phi$, even in the absence of the $\mathbb{Z}_2$ twisted wall.

The only case in which this procedure is potentially sensible is $k=1$. In that case, $Q_\ell$ vanishes, and setting $\mu= 0$ one recovers translational invariance in the $\phi$ direction. In the resulting theory, one can study amplitudes that are proportional to the (infinite) length of $\phi$, like we do in QFT and string theory in flat spacetime. However, this theory is not continuously connected to that with $\mu\not=0$. In particular, in this theory, states with positive and negative $\phi$ momentum are taken to be independent. For any $\mu\not=0$, they are related by reflection from the wall.

A perhaps useful analog is a non-compact orbifold CFT, \eg $\mathbb{R}^n/\mathbb{Z}_2$. This CFT has a continuum of delta-function normalizable states that propagate in the bulk of $\mathbb{R}^n$, and normalizable states localized near the $\mathbb{Z}_2$ fixed point. There are three different theories that one can discuss in this context: (1) the CFT on $\mathbb{R}^n$; (2) the untwisted sector of the CFT on $\mathbb{R}^n/\mathbb{Z}_2$; (3) the full CFT on $\mathbb{R}^n/\mathbb{Z}_2$. Theories (1) and (2) are closely related but are distinct. In particular, while theory (1) is modular invariant, (2) is not, and it requires the twisted sector to become (3), which is modular invariant. And, of course, the CFT (1) is disconnected from (2), (3), in the sense that for general $n$ there is no marginal deformation that connects them.
Physically, theories (1) and (2) differ in the absence of the $\mathbb{Z}_2$ identification in the former. Dropping this identification makes sense if one is studying local scattering processes that occur infinitely far from the fixed point.

The boundary CFT of string theory on $AdS_3\times\mathcal{N}$ with $k=1$ is the analog of theory (3) above. On the other hand, the theory studied in~\cite{Eberhardt:2025sbi} is the analog of theory (1). The restriction of the full string theory on $AdS_3\times{\cal N}$ to the theory of continuous series states with $|w|\ge 1$ is the analog of theory (2). In this sense, the theory of~\cite{Eberhardt:2025sbi} is disconnected from standard string theory on $AdS_3\times{\cal N}$, just like in the case of CFTs (1) and (3).

The discussion above was from the point of view of the boundary theory, the CFT dual to string theory on $AdS_3\times{\cal N}$. It is natural to ask how the above decoupling arises from the bulk perspective.
An interesting question is the following. The symmetric product without the $\mathbb{Z}_2$ twisted wall has a much larger chiral algebra than the full string theory on $AdS_3$. How does one see this symmetry from the bulk perspective? In the next subsections, we answer this question.

\subsection{$\partial_x \phi$ in the fermionic string theory on $AdS_3\times \mathcal{N}$}\label{sec2.3}

For concreteness, we first describe a special case -- the operator $\partial_x\phi$ that is not holomorphic in the full string theory on $AdS_3\times\mathcal{N}$, but becomes holomorphic in its large $\phi$ limit. The mechanism in operation for this operator is similar for other holomorphic operators, that amount to the $\mathbb{R}_\phi$ in the seed of the symmetric product theory, \eqref{sw}.

The breaking of holomorphy for the operator $\partial_x\phi$ is discussed in section 7.2 of \cite{Balthazar:2021xeh}. It is noted there that the operator $\partial_{\bar{x}}\partial_x\phi$ corresponds in the bulk to an operator with $j=1-\frac{k}2$, $m=\bar m=\frac{k}2$. For $k<1$, this operator is non-normalizable, but it sits on an LSZ pole (in the sense of~\cite{Aharony:2004xn}). By deforming away from the pole (see eq. (7.2) in \cite{Balthazar:2021xeh}) and carefully taking the limit, it was shown in \cite{Balthazar:2021xeh} that this operator does not vanish but gives a normalizable operator, (7.5), (7.10), which corresponds in the boundary theory to the $\mathbb{Z}_2$ twisted deformation.

For $k=1$, the analysis of \cite{Balthazar:2021xeh} has an interesting twist (see appendix \ref{appA} for more technical details and facts). In that case, the bulk operator corresponding to  $\partial_{\bar{x}} \partial_x\phi$ has $j=\frac12$, and thus lies at the boundary between the delta-function normalizable and non-normalizable branches. Therefore, one can deform it as in (7.2) of~\cite{Balthazar:2021xeh} in two different ways. One is as done there, which takes this operator into the non-normalizable branch, leading us to study operators of the from $\partial_x\phi\,e^{\beta\phi}$ with real, positive~$\beta$.~\footnote{$\beta=\epsilon Q$ in the notation of eq. (7.2). Note that this involves correcting a typo in \cite{Balthazar:2021xeh}, where in the line above (7.2) $Q$ is replaced by $Q_\ell$.} The other, corresponding to imaginary $\epsilon=ip$, takes it into the delta-function normalizable branch,~\footnote{Note that for the consistency of this procedure it is important that the coefficient of $\epsilon$ in the expression for $m$ in eq. (7.2) of \cite{Balthazar:2021xeh} vanishes for $k=1$. The full expression for $m$ has an extra term, that is quadratic in $\epsilon$, but that is not a problem since it remains real when $\epsilon$ is imaginary.} and to operators of the above form with $\beta=\sqrt{2}ip$.

Approaching $\epsilon\to 0$ from real values leads to the same conclusions as in \cite{Balthazar:2021xeh}:  $\partial_{\bar{x}} \partial_x\phi$ does not vanish, and is given by the $\mathbb{Z}_2$ twisted operator that deforms the symmetric orbifold.
On the other hand, if we approach $\epsilon=0$ from imaginary values (see appendix \ref{appA} for details), we can avoid the LSZ pole, and view the operator $\partial_{\bar{x}} \partial_x\phi$ as a limit of delta-function normalizable operators, for which one can use the construction of \cite{Eberhardt:2025sbi}. However, this comes with a price. If we take $\epsilon=ip$, in the limit $p\to 0$ we are dealing with real $\phi$ momentum that goes to zero. To avoid the LSZ pole, we have to take $\phi\to\infty$ as $p\to 0$. This recovers the conclusion we got from the boundary CFT point of view -- that {\it the decoupled theory of the continuous series states in $AdS_3$ lives at infinite~$\phi$}. It also shows that in the theory describing the infinite $\phi$ region, the operator $\partial_x\phi$ is holomorphic, as one would expect from the boundary perspective.

It is useful to contrast the above discussion with what happens for $k\not=1$. The construction of \cite{Balthazar:2021xeh} allows one to study delta-function normalizable operators in the boundary CFT, of the form $\partial_x\phi\,e^{\beta\phi}$, with $\beta=-\frac{Q_\ell}2+ip$ (eq. (4.7) in \cite{Balthazar:2021xeh}). However, in order to reach the operator $\partial_x\phi$ of interest here, one has to analytically continue this construction to a finite imaginary value of $p$ that corresponds to $\beta=0$. The resulting operator is non-normalizable, and therefore, it does not satisfy the decoupling discussed above.

As mentioned earlier, while the explicit discussion above was for the operator $\partial_x\phi$, it is easy to generalize it to the infinite set of conserved currents of the $\mathbb{R}_\phi$ SCFT that, like $\partial_x\phi$, have the property that they are conserved in the $\phi\to\infty$ theory, but not in the full string theory on  $AdS_3\times{\cal N}$ with $k=1$.
The key point is that, like in the discussion above, for $k=1$ they have $j={1\over 2}$, and can be viewed as limits from the continuum. We will not describe the details here.

To recapitulate, we saw that the theory described in \cite{Eberhardt:2025sbi} is related to the one obtained from the full string theory on $AdS_3$ with $k=1$ by taking the limit $\phi\to\infty$, and focusing on the local physics of delta-function normalizable operators. We also explained how the large chiral algebra of the boundary CFT, that amounts to the $\mathbb{R}$ in the symmetric product with seed $\mathbb{R}\times{\cal N}$, can be obtained in the full $AdS_3$ theory, by studying delta-function normalizable operators in the limit where the $\phi$ momentum goes to zero.

In the following, we will provide some details concerning the limit of delta-function normalizable states, for other chiral operators in $(\mathbb{R}\times{\cal N})^p/S_p$, which also amount to the properties of ${\cal N}$.
A subclass of such currents are conserved in the full theory, and are usually described by operators in the winding-zero sector~\cite{Giveon:1998ns,Kutasov:1999xu}, but one can construct them in the winding-one sector as well.
A subsubclass of examples consists of chiral operators that amount to an affine symmetry of the SCFT ${\cal N}$.
We begin with this example in the next subsection.

\subsection{${\cal K}^a(x)$ in the fermionic string theory on $AdS_3\times{\cal N}$}\label{sec2.4}

Consider the fermionic string on $AdS_3\times{\cal N}$,
and assume that the SCFT on ${\cal N}$ contains a worldsheet supercurrent algebra associated with some group $G$,
generated by worldsheet supercurrents $\chi^a(z)+\theta K^a(z)$.
This implies,~\cite{Giveon:1998ns,Kutasov:1999xu}, that there are spacetime currents corresponding to those in~${\cal N}$.
The physical worldsheet operator in the $w=0$ sector that amounts to the spacetime current ${\cal K}^a(x)$ is
\be\label{ks}
{\cal K}^a(x)\simeq\int d^2z e^{-\varphi-\bar\varphi}\chi^a(z)\bar\psi(\bar x;\bar z)\Phi_{1}(x,\bar x;z,\bar z)
\ee
(see \eg (2.66) in \cite{Balthazar:2021xeh}, and the whole of section 2 there, for a review of prior results and conventions). 

The spacetime currents in~\eqref{ks} satisfy the standard OPE algebra, \cite{Kutasov:1999xu},
\begin{equation}\label{kkope}
\mathcal{K}^a(x)\mathcal{K}^b(y)\sim \frac{\frac{1}{2}\mathcal{I}k_{G}\delta^{ab}}{(x-y)^2}+\frac{f^{ab}_{\,\,\,\,\,\,c}\mathcal{K}^c(y)}{x-y}~,  
\end{equation}
where the operator $\mathcal{I}(x,\bar{x})$ is given by \eg equation (2.68) of \cite{Balthazar:2021xeh}, $k_G$ is the level of the 
worldsheet affine Lie algebra ${\hat G}$ and ${1\over i}f^{ab}_{\,\,\,\,\,\,c}$ are the (real) structure constants of $G$.

Now, for $k=1$,
the operator that amounts to the (same spacetime) current in the $w=1$ sector is (we use the convention $\Phi^{w,j}_h$ for $\bar J=J\equiv h$ in~\cite{Maldacena:2001km}, as in~\cite{Giveon:2023gzh})
\be\label{kwone}
{\cal K}^a(x)\simeq\int d^2z e^{-\varphi-\bar\varphi}\chi^a(z)\bar\psi(\bar x;\bar z)\Phi_{h=1}^{w=1,j=1/2}(x,\bar x;z,\bar z)~,
\ee
where we think about $\Phi_{h=1}^{w=1,j=1/2}(x,\bar x;z,\bar z)$ as
\be\label{lims}
\Phi_{h=1}^{w=1,j=1/2}(x,\bar x;z,\bar z)\equiv\lim_{s\to 0}\Phi_{h=1+s^2}^{w=1,j=1/2+is}(x,\bar x;z,\bar z)~,
\ee
a.k.a. being obtained in the limit of delta-function normalizable operators, when approaching the boundary of the continuum,
namely, by taking the limit where the real $\phi$ momentum goes to zero (see the appendices \ref{appA},\ref{appB} for complementary details and facts).
This way of thinking about $\Phi_{h=1}^{w=1,j=1/2}(x,\bar x;z,\bar z)$ in \eqref{kwone} is required for the operator to live in the infinite $\phi$ regime.

A couple of comments are in order:
\begin{itemize}
\item
The claim above can be proved using~\cite{Kutasov:1999xu} and section~5 in~\cite{Maldacena:2001km}.
\item
The ${\cal K}^a$ in \eqref{ks} and \eqref{kwone} are related, formally, by an FZZ duality; see appendix~\ref{appB}.
\item
It is important to note that while the ${\cal K}^a(x)$ in \eqref{ks} is well defined for {\it any}~$k$,~\cite{Kutasov:1999xu},~\footnote{See also the discussion near (2.66) of~\cite{Balthazar:2021xeh}.}
its FZZ dual cannot be obtained via a limit of delta-function normalizable operators, as in \eqref{lims}, if $k\neq 1$.~\footnote{For example, for $k<1$, the worldsheet operator in (4.34) of~\cite{Balthazar:2021xeh}, which corresponds to ${\cal K}^a_{-1}$,  has $j=1-{k\over 2}$, and is non-normalizable. Note that setting $k=1$ in (4.34) of~\cite{Balthazar:2021xeh} corresponds to the mode ${\cal K}^a_{-1}$ of the ${\cal K}^a(x)$ in~\eqref{kwone}.} 
\item
An interested reader may consult appendix \ref{appA} for related comments.
\end{itemize}

When $k=1$, it is  expected that the currents \eqref{kwone} will satisfy the standard OPE algebra, \eqref{kkope}, where the operator $\mathcal{I}$ in \eqref{kkope} would be the $w=1$ representative of the standard $\mathcal{I}$ operator,
\begin{equation}\label{w=1I}
    \mathcal{I}(x,\bar{x})\simeq\int d^2z  e^{-\varphi-\bar{\varphi}} \psi(x;z)\bar{\psi}(\bar{x};\bar{z})\Phi^{w=1,j=1/2}_{h=1}(x,\bar{x};z,\bar{z})~.
\end{equation}    
To derive the operator product algebra \eqref{kkope} for the $w=1$ presentation of spacetime currents \eqref{kwone}, one needs to use the operator product expansion
\begin{equation}\label{phiope}
    \Phi^{w=1,j=1/2}_{h=1}(x_1,\bar{x}_1;z_1,\bar{z}_1)\Phi^{w=1,j=1/2}_{h=1}(x_2,\bar{x}_2;z_2,\bar{z}_2)\sim \delta^2(x_1-x_2) \Phi^{w=1,j=1/2}_{h=1}(x_2,\bar{x}_2;z_2,\bar{z}_2)~,
\end{equation}
which follows from \cite{Kutasov:1999xu} and FZZ duality (see appendix \ref{appB}).~\footnote{Needless to say, it would be nice to have a direct derivation of \eqref{phiope}.}

The mechanism in operation for ${\cal K}^a(x)$ and $\partial_x\phi$ is similar for other holomorphic operators in the symmetric product theory.
In the next subsections, we'll mention some more examples of such operators in the fermionic string theory on $AdS_3 \times \mathcal{N}$, and in the particular family of the superstring on $AdS_3\times \mathbb{T}^3$.

\subsection{$\mathcal{T}(x)$ in the fermionic string theory on $AdS_3\times\mathcal{N}$}\label{sec2.5}

The spacetime operator $\partial_x \phi$, presented in subsection \ref{sec2.3}, knows only about the $\mathbb{R}_\phi$ in the long-string theory \eqref{sw}, for any $k$, but it is holomorphic only in the limit of delta-function normalizable states
at the large $\phi$, $k=1$ theory, defined in subsections \ref{sec2.2}, \ref{sec2.3} and \ref{sec2.4}. On the other hand, 
the spacetime currents $\mathcal{K}^a(x)$ in the previous subsection are holomorphic for {\it any} $k$, and are blind to the $\mathbb{R}_\phi$  piece in the long-string spacetime seed \eqref{sw}. 

In this subsection, we present a universal example of a spacetime operator -- the $xx$-component of the stress-energy tensor $\mathcal{T}$ -- that is holomorphic for any $k$, yet it's well defined in the winding-one sector of the limit above only when $k=1$, and it knows about the whole seed theory, $\mathbb{R}\times\mathcal{N}$, in this limit.

Concretely, the spacetime CFT$_2$ dual to superstring theory on $AdS_3\times\mathcal{N}$ has a conserved stress-energy tensor, whose holomorphic component is given by (see \eg (2.70) in \cite{Balthazar:2021xeh})
\begin{equation}\label{w=0T}
    \mathcal{T}(x)\simeq\int d^2 z e^{-\varphi-\bar{\varphi}} \left(\partial_x \psi(x;z)\partial_x \Phi_{1}(x,\bar x;z,\bar z)+2\partial_x^2\psi(x;z)\Phi_{1}(x,\bar x;z,\bar z)\right)\bar{\psi}(\bar{x};\bar{z})~,
\end{equation}
where $\Phi_{1}(x,\bar x;z,\bar z)$ is the same non-normalizable vertex operator in $AdS_3$ that features in the definition of the spacetime spin 1 currents $\mathcal{K}^a(x)$ in \eqref{ks}. This stress-energy tensor satisfies the usual OPE algebra, \cite{Kutasov:1999xu},
\begin{equation}\label{ttope}
   \mathcal{T}(x)\mathcal{T}(y)\sim \frac{3k\mathcal{I}}{(x-y)^4}+\frac{2\mathcal{T}(y)}{(x-y)^2}+\frac{\partial_y \mathcal{T}(y)}{x-y}~, 
\end{equation}
where the operator $\mathcal{I}(x,\bar{x})$ is given by \eg eq. (2.68) of \cite{Balthazar:2021xeh}.

When $k=1$, following the same line of arguments as in the case of the conserved spin~1 currents $\mathcal{K}^a(x)$ in subsection \ref{sec2.4}, the $w=1$ representative of the  holomorphic component of the stress tensor is  
\begin{equation}\label{w=1T}
    \mathcal{T}(x)\simeq\int d^2 z e^{-\varphi-\bar{\varphi}} \left(\partial_x \psi(x;z)\partial_x \Phi^{w=1,j=1/2}_{h=1}+2\partial_x^2\psi(x;z)\Phi^{w=1,j=1/2}_{h=1}\right)\bar{\psi}(\bar{x};\bar{z})~.
\end{equation}
Analogous to the affine spacetime currents in their $w=1$ presentation, it is also expected that the above stress-energy tensor component \eqref{w=1T} will satisfy the standard operator product algebra \eqref{ttope} with $k=1$, where the operator $\mathcal{I}$ is given by \eqref{w=1I} (recall that the detailed derivation of the operator product algebra \eqref{ttope} requires the operator product expansion \eqref{phiope}).
 
\subsection{The superstring on $AdS_3\times{\cal N}$ with $k=1$}\label{sec2.6}

Superstrings on $AdS_3\times{\cal N}$  that give rise to $(2,2)$ superconformal spacetime CFT were constructed in~\cite{Giveon:1999jg,Berenstein:1999gj,Giveon:2003ku}.
The main ingredient of the construction is the following. 
If $\cal N$ has an affine $U(1)_L\times U(1)_R$ symmetry and ${\cal N}/U(1)$ has a $(2,2)$ worldsheet superconformal symmetry, then there is a chiral GSO construction of a superstring with two-dimensional $(2,2)$ spacetime superconformal symmetry, in which the worldsheet $U(1)_L\times U(1)_R$ gives rise to the $R$-symmetry of the spacetime supersymmetry.
String theory on $AdS_3\times \mathbb{T}^3$  allows such a chiral GSO construction, giving rise to a family of superstrings on $AdS_3$ with $k=1$
(since eq. \eqref{k} implies that the total level $k$ of $AdS_3$ is
\be\label{kone}
k=1~,
\ee
in the case that ${\cal N}$ is the SCFT on $\mathbb{T}^3$).

In this subsection, we comment on superstring theory on $AdS_3\times \mathbb{T}^3$: 
\begin{itemize}
\item
String theory on $AdS_3\times \mathbb{T}^3$ is obtained by studying a system of intersecting NS fivebranes (or its T-dual conifold) and fundamental strings (see \eg \cite{Giveon:2005mi} and refs. therein). The resulting background contains both discrete (normalizable) and continuous (delta-function normalizable) states, which interact with each other. This is the general situation in $AdS_3/CFT_2$, for any $k$, including $k=1$ (as discussed above). 
\item
As discussed above, in the construction of~\cite{Eberhardt:2025sbi}, and/or our $k=1$ limit,
one discards the normalizable states and drops the identification between delta-function normalizable states with positive and negative radial momenta. In that sense, it is disconnected from the standard construction of the item above, just like CFT on $\mathbb{R}^n$ is disconnected from that on $\mathbb{R}^n/\mathbb{Z}_2$.
In the following items, we restrict to the properties of the superstrings on $AdS_3\times \mathbb{T}^3$ in this limit.
\item
The construction of the spacetime $U(1)$ $R$-currents (from the left and right worldsheet $U(1)$ supercurrents associated with them) is a particular case of the ${\cal K}^a(x)$ (and their antiholomorphic analogs, $\bar{\cal K}^a(\bar x)$) in subsection \ref{sec2.4}. But, there is an enhancement of the $R$-symmetry in the limit, which we mention next.
\item
On top of that, one can repeat in 1--1 everything that was done in~\cite{Balthazar:2021xeh}.~\footnote{Not only for the type II case, but also to the type 0 one.}
But, there will be an enhancement of chiral symmetry for {\it any} $\mathbb{T}^3$ -- the symmetry of SCFT on $\mathbb{R}\times \mathbb{T}^3$.
E.g. the affine left and right $U(1)$ $R$-symmetries in the previous item are enhanced to affine $SU(2)$ ones, 
similar to the way that $\partial_x\phi$ becomes holomorphic in the limit. 
Again, the key point is that, like in the discussion above, for $k=1$ the worldsheet operators in the $w=-1$ sector that amount to the holomorphic operators in the long-string theory, $\mathbb{R}\times \mathbb{T}^3$,  have $j={1\over 2}$, and can be viewed as limits from the continuum.
We leave the details as an exercise to the reader.
\item
Needless to say that, in particular, for the superstring, in addition to the enhancement of the left and right $U(1)$ $R$-currents above (${\cal K}(x)$ and $\bar{\cal K}(\bar x)$) to $SU(2)$ ones, and the appearance of new left and right $\mathbb{R}_\phi$ currents ($\partial_x\phi$ and $\partial_{\bar x}\phi$), in the large~$\phi$ decoupling limit of the $k=1$ theory, we'll find enhancement of the $(2,2)$ spacetime supersymmetry currents corresponding to on-shell worldsheet operators in the winding-one sector (which can be constructed along the lines of~\cite{Balthazar:2021xeh} also when $k=1$) to a set generating a $(4,4)$ superconformal symmetry (along the lines above), for {\it any}~$\mathbb{T}^3$.
Again, we'll not present the details here, apart for an example in the next item.
\item 
As an example, consider the following.
In the type II superstring on \be\label{adsst}
AdS_3\times\mathbb{S}^1\times\mathbb{T}^2~,
\ee
with $N=(2,2)$ spacetime supersymmetry constructed as in~\cite{Giveon:1999jg,Berenstein:1999gj,Giveon:2003ku} w.r.t. the $U(1)_L\times U(1)_R$ of the $\mathbb{S}^1$ in \eqref{adsst}, the decoupled long-strings symmetric product SCFT has in its seed a non-compact free boson $\phi$. Since $\phi$ is free, the current $\partial_x\phi$ in the block of the symmetric product, $\mathbb{R}\times\mathbb{S}^1\times\mathbb{T}^2$, is holomorphic and conserved, $\partial_{\bar{x}}\partial_x \phi=0$ (see subsection \ref{sec2.3} and appendix \ref{appA} for more details).
Its worldsheet dual is given by the on-shell vertex operator \eqref{aa}.
In the $(2,2)$ superstring on \eqref{adsst}, the holomorphic current $\partial_x\phi$ has lower $N=2$ superfield components -- holomorphic complex fermions, $\mathcal{S}^\pm$, which are dimension  $\left(\frac{1}{2},0\right)$ operators in the seed of the symmetric-product SCFT, $(\mathbb{R}\times\mathbb{S}^1\times\mathbb{T}^2)^p/S_p$. The on-shell worldsheet vertex operators corresponding to them are obtained following a similar procedure to that leading to (6.15) of \cite{Balthazar:2021xeh}, by performing the $s\to 0$ limit of $j={1\over 2}+is$ in (6.14), similar to what is done for $\partial_x\phi$ in subsection \ref{sec2.3} and appendix \ref{appA}. The result is
\begin{equation}\label{ss}
    \mathcal{S}^\pm\longleftrightarrow\ e^{-\frac{\varphi}{2}-\bar{\varphi}}e^{\pm \frac{i}{2}Z\pm i \frac{Y}{\sqrt{2}}}e^{\frac{3}{2}iH_{sl}+i\bar{H}_{sl}\pm \frac{i}{2}H_3}\Phi^{(-1)}_{{1\over 2};-{1\over 2},{1\over 2}}~,\qquad\partial_{\bar x}\partial_x\mathcal{S}^\pm=0~,
\end{equation}
where $Y$ and $Z$ are canonically normalized, the former corresponding to the $\mathbb{S}^1$ in \eqref{adsst} and the latter is the bosonization of the $U(1)$ $R$-currents of the $N=(2,2)$ SCFT on the $\mathbb{T}^2$ in \eqref{adsst}.
\item
Finally, note that the superstring on $AdS_3\times \mathbb{T}^3$ includes as a special case the string theory on $AdS_3\times \mathbb{S}^3\times \mathbb{S}^3\times \mathbb{S}^1$ with $k=1$ (at a particular point in the Narain moduli space of the $\mathbb{T}^2$ in \eqref{adsst}, where the SCFT on $\mathbb{T}^2$ is equivalent to an $SU(2)_2\times SU(2)_2$ SCFT, that is equivalent to six free fermions), which has a large $(4,4)$ supersymmetry in spacetime,~\cite{Elitzur:1998mm}.
\end{itemize}

\subsection{Single-trace $T\bar T$ and $J\bar T$ in $AdS_3\times{\cal N}$ with $k=1$}\label{sec2.7}

The spacetime theory dual to string theory on $AdS_3\times \mathcal{N}$ has certain single-trace irrelevant operators, which have been the subject of contemporary interest in relation to understanding holography in spacetime beyond asymptotically AdS and solvable irrelevant deformations of two-dimensional CFTs 
(see \eg \cite{Giveon:2017nie,Jiang:2019epa,Apolo:2019zai,Chakraborty:2019mdf,Apolo:2021wcn,Giveon:2024sgz} and references therein). 
These operators were first constructed in \cite{Kutasov:1999xu}, and are often referred to in the literature as single-trace $T\bar{T}$ and single-trace $J\bar{T}$ operators. In this subsection, we will review these irrelevant operators and present their $w=1$ form in the $k=1$ limit.

The single-trace $T\bar{T}$ operator, often denoted by $D(x,\bar{x})$, is a dimension $(2,2)$ quasi-primary operator of the spacetime Virasoro algebra,  which in the fermionic string on $AdS_3$ is given by
\begin{equation}\label{D}
    D(x,\bar{x})\simeq \int d^2z e^{-\varphi-\bar{\varphi}} \left(\partial_x\psi(x;z)\partial_x+2\partial_x^2\psi(x;z)\right)\left(\partial_{\bar{x}}\bar{\psi}(\bar{x};\bar{z})\partial_{\bar{x}}+2\partial_{\bar{x}}^2\bar{\psi}(\bar{x};\bar{z})\right)\Phi_{1}(x,\bar{x};z,\bar{z})~.
\end{equation}
The operator $D(x,\bar{x})$ is single-trace because it scales like $p\sim 1/g_s^2$, where $g_s$ is the string coupling in $AdS_3$, as opposed to the double-trace ${\cal T}(x)\bar{\cal T}(\bar{x})$ which scales like $p^2\sim 1/g_s^4$.
Thus, the single-trace operator $D(x,\bar{x})$ and the double-trace operator ${\cal T}(x)\bar{\cal T}(\bar{x})$ are two distinct operators, despite the fact that they have the same OPEs with ${\cal T}(x)$ and~$\bar{\cal T}(\bar x)$:
the former is a massive mode of the dilaton-graviton sector of string theory on~$AdS_3$, which in particular can be added to the worldsheet Lagrangian and deform $AdS_3$, as we'll review in the next section.

When $k=1$, one can easily write down, along the lines in the previous subsections, the $w=1$ representative of the operator $D(x,\bar x)$,
\begin{equation}\label{D1}
    D(x,\bar{x})\simeq\int d^2z e^{-\varphi-\bar{\varphi}} \left(\partial_x\psi(x;z)\partial_x+2\partial_x^2\psi(x;z)\right)\left(\partial_{\bar{x}}\bar{\psi}(\bar{x};\bar{z})\partial_{\bar{x}}+2\partial_{\bar{x}}^2\bar{\psi}(\bar{x};\bar{z})\right)\Phi_{h=1}^{w=1,j=1/2}~.
\end{equation}
Recall that in the decoupled long-strings theory, obtained by setting the coefficient of the marginal $\mathbb{Z}_2$-wall to zero, the spacetime theory is a symmetric product,  $(\mathbb{R}\times \mathcal{N})^p/S_p$. 
In this decoupled theory, the operator in \eqref{D1} can be thought of as the single-trace $T\bar{T}$ operator of the symmetric product,
\begin{equation}
    D(x,\bar{x}) \simeq \sum_{i=1}^p {\cal T}_i(x)\bar{\cal T}_i(\bar{x})~,
\end{equation}
where ${\cal T}_i$ $(\bar{\cal T}_i)$ is the holomorphic (anti-holomorphic) component of the stress-energy tensor operator of the $i^{th}$ block of the symmetric product SCFT, $(\mathbb{R}\times \mathcal{N})^p/S_p$.

As pointed out in \cite{Kutasov:1999xu}, string theory on $AdS_3\times\mathcal{N}$, for general $k$, also possesses a spin-1 irrelevant single-trace operator, $A^a(x,\bar{x})$, which is a dimension $(1,2)$ quasi-primary operator of the spacetime Virasoro algebra, if ${\cal N}$ has an affine symmetry generated by $K^a(z)$. 
This operator is often termed as the single-trace $J\bar{T}$ operator. 
In the fermionic string on $AdS_3\times \mathcal{N}$, 
if the SCFT on ${\cal N}$ contains a worldsheet supercurrent algebra associated with some group $G$, generated by worldsheet
supercurrents $\chi^a(z)+\theta K^a(z)$,
$A^a(x,\bar{x})$ is given by
 \begin{equation}\label{A}
     A^a(x,\bar{x})=\int d^2z e^{-\varphi-\bar{\varphi}} \chi^a(z)\left(\partial_{\bar{x}}\bar{\psi}(\bar{x};\bar{z})\partial_{\bar{x}}+2\partial^2_{\bar{x}}\bar{\psi}(\bar{x};\bar{z})\right) \Phi_1(x,\bar{x};z,\bar{z})~.
 \end{equation}
As above, this operator is distinct from its double-trace ${\cal K}^a(x)\bar{\cal T}(\bar{x})$ cousin, despite sharing a set of the same quantum numbers.

In the $k=1$ limit above, the $w=1$ representative of $A^a(x,\bar{x})$ is given by
\begin{equation}\label{A1}
    A^a(x,\bar{x})=\int d^2z e^{-\varphi-\bar{\varphi}} \chi^a(z)\left(\partial_{\bar{x}}\bar{\psi}(\bar{x};\bar{z})\partial_{\bar{x}}+2\partial^2_{\bar{x}}\bar{\psi}(\bar{x};\bar{z})\right) \Phi_{h=1}^{w=1,j=\frac{1}{2}}~.
\end{equation}
This operator can be thought of as the single-trace $J\bar{T}$ operator of the symmetric product SCFT, $(\mathbb{R}\times \mathcal{N})^p/S_p$, 
\begin{equation}
   A^a(x,\bar{x}) \simeq \sum_{i=1}^p {\cal K}^a_i(x)\bar{\cal T}_i(\bar{x})~,
\end{equation}
in the $k=1$ decoupling limit of the long-strings theory.

In the following sections, we will inspect the single-trace $T\bar{T}$ deformation of the fermionic string on $AdS_3\times \mathcal{N}$, 
with an emphasis on the superstring on deformed $AdS_3\times\mathbb{T}^3$,
and discuss certain aspects of its single-trace $J\bar{T}$ deformation. 

\section{String theory on ${\cal M}_3\times{\cal N}$ and on ${\cal M}_3\times\mathbb{T}^3$}\label{sec3}

In this section, after reviewing some properties of single-trace $T\bar T$ holography (in subsection \ref{sec3.1}),
we inspect the superstring on the deformed $AdS_3$ sigma-model, ${\cal M}_3$, at $k=1$, and mention properties of the decoupled theory in its asymptotically linear dilaton flat spacetime, $\mathbb{R}^{1,1}\times \mathbb{R}_\phi$ (and/or $\mathbb{R}_t\times \mathbb{S}^1\times \mathbb{R}_\phi$), in particular, its predictions for $T\bar T$-deformed $\mathbb{R}\times \mathbb{T}^3$ SCFT (in subsection~\ref{sec3.2}).
As a couple of examples, we then obtain the $T\bar T$-deformed 2pf of ${\cal K}^a(x)$ (in subsection \ref{sec3.3}), and of ${\cal T}(x)$ (in subsection~\ref{sec3.4}).

\subsection{${\cal M}_3$ and single-trace $T\bar T$ holography -- a review}\label{sec3.1}

We begin in this subsection by recalling some properties of deformed $AdS_3$ and single-trace $T\bar T$ holography, that will be useful later.

Single-trace $T\bar T$ holography~\cite{Giveon:2017nie} is obtained from the $AdS_3/CFT_2$ one by adding to the Lagrangian of the spacetime CFT$_2$  the dimension $(2,2)$ quasi-primary operator $D({\vec x})\equiv D(x,\bar x)$, ${\vec x}\equiv(x^0,x^1)$, constructed in~\cite{Kutasov:1999xu} and reviewed in subsection \ref{sec2.7}.
In the fermionic string, it amounts on the worldsheet to the on-shell operator in \eqref{D}.
This irrelevant deformation can be controlled, since in the dual bulk string theory on $AdS_3$ it corresponds to a current-current deformation of the $SL(2,\mathbb{R})$ WZW model, by  $\lambda J^-\bar J^-$,~\cite{Giveon:2017nie}.

The Lagrangian of the deformed $SL(2,\mathbb{R})$ worldsheet CFT is
\be\label{l}
{\cal L}=k\partial\phi\bar\partial\phi+{\lambda\over\lambda+e^{-2\phi}}\partial x^+\bar\partial x^-~,
\ee
with a dilaton
\be\label{dilaton}
\Phi={\rm const}-\log\left(1+\lambda e^{2\phi}\right)~.
\ee
The geometry of this sigma-model background, ${\cal M}_3$, thus interpolates between a flat spacetime,
\be\label{asymp}
\mathbb{R}^{1,1}\times \mathbb{R}_\phi~,
\ee
with a linear dilaton, and a (Poincar\'e patch of)   $AdS_3$ cap in the IR, $\phi\to -\infty$ (where it's natural to rescale $x^\pm$ by a factor of $\sqrt{\lambda/k}$,~\cite{Giveon:2017myj}).

A useful description of the deformed worldsheet theory \eqref{l} is obtained by a null gauging of the SCFT
\be\label{rads}
\mathbb{R}^{1,1}\times AdS_3~,
\ee
in the directions
\be\label{null}
i\partial(y-t)+\alpha J^-~,\qquad i\bar\partial(y+t)\pm\alpha \bar J^-~,
\ee
where $(t,y)$ are the canonically normalized bosonic fields of the $\mathbb{R}^{1,1}$ in \eqref{rads},
and the $\alpha$ in \eqref{null} is related to the $\lambda$ in~\eqref{l} via $\lambda=\pm\alpha^2$, with the $\pm$ amounting to axial and vector gauging in \eqref{null}, respectively.
The resulting fermionic string theory on ${\cal M}_3\times{\cal N}$ is now obtained along the lines presented in~\cite{Giveon:2017myj,Asrat:2017tzd}.

Following~\cite{Asrat:2017tzd},
a large class of observables in the fermionic string on ${\cal M}_3\times{\cal N}$ is given by
vertex operators in the (NS,NS) sector, which take the form (in the $(-1,-1)$ picture)
\be\label{o}
O^{w,j}({\vec p})\simeq\int d^2z e^{-\varphi-\bar\varphi}\Phi_{h}^{w,j}({\vec p})e^{-i\alpha(p_0 t+p_1 y)}O_{\cal N}~.
\ee
Here $\varphi,\bar\varphi$ are worldsheet fields associated with the superconformal ghosts, that keep track of the picture, and
$\Phi_{h}^{w,j}({\vec p})$ are vertex operators on $AdS_3$ in the winding $w\geq 0$ sectors,
obtained by spectral flow from operators in the representation $j$ of $SL(2,\mathbb{R})$;
they are the Fourier transforms of the operators $\Phi_h^{w,j}(x,\bar x;z,\bar z)$ in~\cite{Maldacena:2001km} to momentum space,
${\vec p}\equiv(p_0,p_1)$.~\footnote{Recall that
in the $(x,\bar x)$ basis, we can label the operators with $w\geq 0$; see below (5.13) in~\cite{Maldacena:2001km}.
In~\cite{Maldacena:2001km}, $h$ was denoted by $J$, and we consider operators with $\bar J=J\equiv h$.}

The operators in eq. \eqref{o} include, in addition to the set of operators in the $w=0$ sector (in which case $h=j$),
considered in eq. (2.9) of~\cite{Asrat:2017tzd} (in which case the $AdS_3$ operator is $\Phi_h^{w=0,j=h}(x,\bar x;z,\bar z)\equiv\Phi_h(x,\bar x;z,\bar z)$),
also those with non-zero winding, $w>0$
(in which case the spacetime conformal weight $h$ of the $AdS_3$ operator $\Phi_h^{w>0,j}(x,\bar x;z,\bar z)$ and its $j$ are different quantum numbers).
When $w>0$, the mass-shell condition for the operators in eq.~\eqref{o} is~\footnote{This is obtained using \eg eq. (4.6) in~\cite{Giveon:2019fgr},
which follows from eqs. (3.4),(3.5) in~\cite{Maldacena:2001km};
we take $\bar\Delta_{\cal N}=\Delta_{\cal N}$, so that $\bar h=h$ below, as in~\cite{Asrat:2017tzd}.}
\be\label{onshell}
-{j(j-1)\over k}-w\left(h_{p^2}-{k\over 4}w\right)+{\alpha'\over 4}\lambda p^2+\Delta_{\cal N}={1\over 2}~.
\ee
The subscript $p^2$ was added to $h$, to emphasize that it depends on $p^2\equiv{\vec p}^{\,2}$ via the on-shell condition.
Denoting by $h$ the value of $h_{p^2}$ in the undeformed theory, 
\be
h\equiv\lim_{{\vec p}\to 0}h_{p^2}~,
\ee
from eq.~\eqref{onshell} one finds that, for $w\neq 0$,
\be\label{hp}
h_{p^2}=h+{\alpha'\over 4}{\lambda\over w} p^2~,
\ee
for fixed $j={1\over 2}+is$ in~\eqref{onshell}, as needed for a $T\bar T$ trajectory~\cite{Giveon:2017myj,Giveon:2019fgr} (see below). \footnote{Equation \eqref{hp} was also obtained by a different method in \cite{Cui:2023jrb}.}

Compactifying $y$ in \eqref{o} on a circle with radius $R$ (which gives rise to compactifying the $x^1$ of ${\cal M}_3$ in \eqref{l},~\cite{Giveon:2017myj}), as well as defining $E$ to be the energy relative to a BPS string wrapping the circle
(which corresponds to the supersymmetric vacuum), and fixing the gauge in the $AdS_3$ factor of \eqref{rads},~\footnote{As opposed to the fixing above, which was done in the $\mathbb{R}^{1,1}$ factor, instead.} a.k.a. restricting to $\Phi_h^{w=0,j=h}({\vec p})\equiv\Phi_h({\vec p})$ in \eqref{o} and replacing
\be\label{expy}
e^{-i\alpha(p_0 t+p_1 y)}\to e^{-i\left(E+{wR\over\alpha'}\right)t}e^{ip_L y+ip_R y}~,
\ee
with
\be\label{plpr}
(p_L,p_R)=\left({n\over R}+{wR\over\alpha'},{n\over R}-{wR\over\alpha'}\right)~,
\ee
one finds that the mass-shell condition, for fixed $j={1\over 2}+is$, leads to the spectrum of
a symmetric-product theory whose seed is the $T\bar T$-deformed long-string theory;~\footnote{Alternatively, one can obtain the same results by a null gauging of $\mathbb{R}_t\times \mathbb{S}^1_y\times AdS_3$.} see section 3 in~\cite{Giveon:2017myj} for details.

To recapitulate, we recalled in this subsection the facts needed to address straightforwardly the special case of interest, $k=1$, in the rest of this section.

\subsection{The superstring on ${\cal M}_3\times \mathbb{T}^3$ and $T\bar T$-deformed $\mathbb{R}\times \mathbb{T}^3$}\label{sec3.2}

We are now ready to turn to the decoupled theory in the asymptotically linear dilaton flat spacetime of ${\cal M}_3$,~\eqref{asymp} and/or its compactification on a circle,  which can be obtained by a consistent limit in the special $k=1$ case.
As before, we focus on the superstring on ${\cal M}_3\times \mathbb{T}^3$ (just for concreteness of the presentation).

From the bulk point of view, it follows from the review in the previous subsection that the arguments and manipulations for the fermionic string on $AdS_3\times{\cal N}$ with $k=1$ propagate straightforwardly to its deformed fermionic string on ${\cal M}_3\times{\cal N}$.
From the point of view of the two-dimensional holographic dual, it's obvious that one obtains a symmetric product of a $T\bar T$-deformed $\mathbb{R}\times{\cal N}$ in the decoupled theory at $\phi\to\infty$.  

For the superstring on ${\cal M}_3\times \mathbb{T}^3$, the boundary theory is thus a symmetric product whose seed is a $T\bar T$-deformed $(4,4)$ supersymmetric $\mathbb{R}\times \mathbb{T}^3$ CFT.
Restricting to the winding-one sector in this superstring theory, one can thus predict precise properties of the $T\bar T$-deformed $\mathbb{R}\times \mathbb{T}^3$ SCFT.
In particular, the deformed properties of the currents corresponding to the enhanced $(4,4)$ supersymmetry can be followed.
Namely, after repeating in 1--1 everything that was done in~\cite{Balthazar:2021xeh},
with the addition of the enhancement of chiral symmetry for {\it any} $\mathbb{T}^3$ -- the symmetry of SCFT on $\mathbb{R}\times \mathbb{T}^3$,
one should deform the manipulations and interpretations along the lines above.
Again, the key point is that, like in the discussion above, for $k=1$ the worldsheet operators in the $w=-1$ sector that amount to the holomorphic operators in the long-string theory,  have $j={1\over 2}$, and can be viewed as limits from the continuum; we leave the details to future work.

In the next subsections, we present a couple of examples of such predictions.

\subsection{Deformed 2pf of ${\cal K}^a(x)$ in $T\bar T$-deformed $\mathbb{R}\times \mathbb{T}^3$}\label{sec3.3}

The superstring on $AdS_3\times \mathbb{T}^3$ is a particular case of the fermionic string on $AdS_3\times{\cal N}$ with $k=1$, where ${\cal N}$ has affine left and right $U(1)$ symmetries, with holomorphic and antiholomorphic currents $K^a(z)$ and ${\bar K}^a(\bar z)$, due to the SCFT on $\mathbb{T}^3$. Correspondingly, the boundary theory has affine ${\cal K}^a(x)$ and $\bar{\cal K}^a(\bar x)$ currents, which amount to worldsheet operators in the winding one sector,~\eqref{kwone} with~\eqref{lims} (and their right-handed analogs).

From the discussion above, one can thus obtain the $T\bar T$-deformed 2pf of ${\cal K}^a(x)$ straightforwardly,
following a minor generalization of~\cite{Giveon:2023gzh} (whose conventions we used),~\footnote{Comments and typos to~\cite{Giveon:2023gzh}:
\begin{enumerate}
\item
Equation (7) in~\cite{Giveon:2023gzh} is obtained for fixed $j$ in (5) (as needed for a $T\bar T$ trajectory,~\cite{Giveon:2017myj,Giveon:2019fgr}).
\item
The ${kw\over 4}$ in (5) of~\cite{Giveon:2023gzh} should be ${k+2\over 4}w$.
\item
The $2h+1$ in (8) of~\cite{Giveon:2023gzh} should be $2h-1$.
\end{enumerate}
} 
via the steps:
\begin{enumerate}
\item
Start with the 2pf of $\Phi_{h=1}^{w=1,j=1/2}(x,\bar x;z,\bar z)$ (defined in the limit~\eqref{lims}) from~\cite{Maldacena:2001km},
which is $\sim(h{\rm-dependent}\,\,{\rm factor})|x|^{-4h}$.
\item
Multiply it by the $\sim\bar x^2$ from the 2pf of $\bar{\psi}(\bar x;\bar z)$ (see (8.2) in~\cite{Kutasov:1999xu}).
\item 
Fourier transform it (from $(x,\bar x)$ space to $(p,\bar p)$ momentum space). 
\item
Replace the $h$ dependence from~\cite{Maldacena:2001km}, wherever it appeared (prior to setting it to~$1$),
with $h_{p^2}=h+{\alpha'\over 4}\lambda p^2$, \eqref{hp}, where $p^2\equiv p\bar p$, as in~\cite{Giveon:2023gzh}.
\item
Recall that we regard states with positive and negative $\phi$ momentum (a.k.a. with $s>0$ in $j={1\over 2}+is$ and $-s$, respectively, prior to taking the limit~\eqref{lims} above) as being independent in the strict $\phi\to\infty$ limit, and thus ignore the reflection coefficient in the outcome (a.k.a. the phase $R(p^2)=e^{i\delta(p^2)}$ in (9) with (10) of~\cite{Giveon:2023gzh}).
\item
Finally, set $h=1$ everywhere it appears. 
\end{enumerate}
The result thus obtained is 
\be\label{kk}
\langle\mathcal{K}^a(p,\bar p)\mathcal{K}^b(-p,-\bar p)\rangle_{\rm target}=\delta^{ab}\,{\bar p\over p}\,{\Gamma\left(-1-{\alpha'\over 2}\lambda p\bar p\right)\over\Gamma\left({\alpha'\over 2}\lambda p\bar p\right)}\left({p\bar p\over 4}\right)^{{\alpha'\over 2}\lambda p\bar p}~.
\ee 
Note that we normalized the two-point function such that the r.h.s. of~\eqref{kk} has the form $\delta^{ab}\,{{\bar p}\over p}\left(a(p\bar p/\mu^2)+1\right)$, with $a(p\bar p/\mu^2)\to 0$ when $p\bar p/\mu^2\to 0$ and the~$1$ corresponding to the current algebra level in the IR, as it should on general grounds.
The equality in eq.~\eqref{kk} was thus set by restricting to the winding-one long-string sector.

To recapitulate, the deformed momentum-space 2pf of the (undeformed) $\mathcal{K}^a$, defined in \eqref{kwone} with \eqref{lims}, in the decoupled long-string spacetime theory of string theory on $\mathcal{M}_3\times\mathbb{T}^3$, is~\eqref{kk}. 
String theory thus predicts such a 2pf, \eqref{kk}, in $T\bar T$-deformed $\mathbb{R}\times\mathbb{T}^3$ SCFT. 

In the next subsection, we'll provide a similar prediction for the deformed 2pf of the ($xx$-component of the) stress-energy tensor in the $\mathbb{R}\times\mathcal{N}$ seed of the boundary theory.

\subsection{Deformed 2pf of $\mathcal{T}(x)$ in $T\bar T$-deformed $\mathbb{R}\times\mathcal{N}$ with $k=1$}\label{sec3.4}

For the $xx$-component of the stress-energy tensor of the boundary theory, the $\mathcal{T}(x)$ in the undeformed CFT of subsection~\ref{sec2.5},~\eqref{w=1T}, defined in the limit \eqref{lims}, a similar calculation to the one leading to \eqref{kk}, 
gives the momentum-space deformed 2pf of $\mathcal{T}(x)$, via the steps:
\begin{enumerate}
\item Start with the $w=1$ presentation of the stress-energy tensor in~\eqref{w=1T}, with $\Phi_{h=1}^{w=1,j=1/2}$ defined in the limit \eqref{lims}, and consider its 2pf.
\item 
Using the properties of $\psi,\bar\psi$ (see \cite{Kutasov:1999xu}), and those of $\Phi^{w=1,j=1/2+\epsilon}_{h=1-\epsilon^2}$ (see subsection \ref{sec2.4} and appendix \ref{appB}, with the properties of $\Phi_h$ in \cite{Kutasov:1999xu}) continued to imaginary $\epsilon$ ($\epsilon=is$, as in appendix \ref{appA}, in the $s\to 0$ limit), one finds that
the only term that contributes to $\langle\mathcal{T}(x)\mathcal{T}(0)\rangle$ is 
\begin{equation}\label{phiphia}
\lim_{s\to 0}\langle[\partial_x\psi\partial_x\Phi^{w=1,j=1/2+is}_{h=1+s^2}\bar{\psi}](x,\bar x)[\partial_y\psi\partial_y \Phi^{w=1,j=1/2+is}_{h=1+s^2}\bar{\psi}](y,\bar y)\rangle|_{y=0}\simeq\lim_{s\to 0} h(1+2h)\frac{\bar{x}^4}{(x\bar{x})^{2(h+1)}}~.
\end{equation}
\item 
Since we are interested  in the long-strings decoupling limit of the fermionic string on $AdS_3\times\mathcal{N}$ at $k=1$,
we ignore the reflection coefficient in the two-point function of $\Phi^{w=1,j=1/2+is}_{h=1+s^2}$ (see the discussion in item 5 of subsection ~\ref{sec3.3}). 
\item 
Next, we consider the Fourier transform of the r.h.s. of \eqref{phiphia},
\begin{equation}\label{fta}
h(1+2h) \int d^2x~ e^{i(p\bar{x}+\bar{p}x)}\frac{\bar{x}^4}{(x\bar{x})^{2(h+1)}}\simeq \frac{\Gamma(3-2h)}{\Gamma(2h)}\frac{\bar{p}^2}{p^2}\left(\frac{p\bar{p}}{4}\right)^{2h-1}~.
\end{equation}
\item 
Finally, we replace $h\to h_{p^2}=1+\frac{\alpha'}{4}\lambda p\bar{p}$ in \eqref{fta}, and normalize the correlation function such that it gives the standard 2pf in the undeformed CFT limit of the spacetime seed, namely, 
\begin{equation}\label{TTcfta}
\lim_{\lambda\to 0} \langle\mathcal{T}(p,\bar p)\mathcal{T}(-p,-\bar p)\rangle_{\rm target}=6\frac{\bar{p}^3}{p}~,  
\end{equation}
where the $6$ on the r.h.s. is the central charge of the $\mathbb{R}\times\mathcal{N}$ seed SCFT. 
\end{enumerate}
The resulting momentum-space deformed 2pf, thus obtained, is
\be\label{tt}
\langle\mathcal{T}(p,\bar p)\mathcal{T}(-p,-\bar p)\rangle_{\rm target}=6{{\bar p}^3\over p}\,{\Gamma\left(-1-{\alpha'\over 2}\lambda p\bar p\right)\over\Gamma\left({\alpha'\over 2}\lambda p\bar p\right)}\left({p\bar p\over 4}\right)^{{\alpha'\over 2}\lambda p\bar p}~.
\ee
The equality in \eqref{tt} amounts to restricting to the winding-one long-string sector.

To recapitulate, eq.~\eqref{tt} provides another prediction for $T\bar T$-deformed SCFT on $\mathbb{R}\times{\cal N}$.
Note that on general grounds, we should have obtained ${{\bar p}^3\over p}\left(f(p\bar p/\mu^2)+c\right)$ on the r.h.s. of \eqref{tt}, with $f(p\bar p/\mu^2)\to 0$ when $p\bar p/\mu^2\to 0$ and $c$ being the central charge of the undeformed CFT, 
which is indeed the case.~\footnote{It is intriguing to note though that in the $T\bar T$-deformed string theory on $AdS_3\times\mathbb{T}^3$, ${\bar p}^2\langle\mathcal{K}^a(p,\bar p)\mathcal{K}^b(-p,-\bar p)\rangle_{\rm target}={1\over 6}\delta^{ab}\langle\mathcal{T}(p,\bar p)\mathcal{T}(-p,-\bar p)\rangle_{\rm target}$, \eqref{kk}, \eqref{tt}. 
It's likely related to the fact that the stress-energy tensor of the undeformed SCFT on $\mathbb{R}\times\mathbb{T}^3$ is obtained from its affine supercurrents via the Sugawara construction.}

\section{Discussion}\label{diss}

In this note, we revisited the fermionic string theory on $AdS_3\times\mathcal{N}$ with $k=1$, and considered its single-trace $T\bar{T}$ deformation. 
As a concrete family of examples, we focused on the $(2,2)$ superstring on (deformed) $AdS_3\times{\mathbb T}^3$. 
The new string theory on $AdS_3\times\mathcal{N}$ at $k=1$ ($k=3$ for the bosonic case), defined in \cite{Eberhardt:2025sbi}, whose holographic dual is the $p$-fold symmetric product of the long-strings theory, $(\mathbb{R}\times\mathcal{N})^p/S_p$, can be obtained from the point of view of the boundary theory  by setting to zero the coefficient of the $\mathbb{Z}_2$-twisted marginal wall in the construction of \cite{Chakraborty:2025nlb}.
From the bulk point of view, we argued that this is equivalent to a certain limiting procedure, which is valid only for $k=1$.
In this $k=1$ limit, one is left with a decoupled theory of delta-function normalizable states of affine $SL(2,\mathbb{R})$ in {\it non}-zero winding sectors (while discarding both normalizable and non-normalizable operators in the theory), and regarding (real) positive and negative radial momenta as independent. 

This may seem puzzling because, as per the standard holographic dictionary, the non-normalizable operators on the gravity side correspond to local operators in the dual boundary theory.~\footnote{And since positive and negative real radial momenta amount to an incoming and outgoing reflected waves, respectively, and thus are dependent; in appendix \ref{appC} we'll illustrate a special property of the phase shift at $k\to 1$, which nevertheless seems in harmony with our decoupled $k=1$ theory.} In this paper, we argue that such non-normalizable operators in string theory on $AdS_3 \times \mathcal{N}$ at $k=1$ have local representatives in the winding-one sector of the delta-function normalizable branch. In particular, we inspected the winding-one representatives of conserved currents, \eg the spacetime affine currents and the stress-energy tensor operator, and argued that they satisfy the standard OPE algebra. In the long-strings decoupling limit, there are additional emergent spin-1 conserved currents, \eg $\partial_x\phi$ and its $N=2$ fermionic superpartners, $\mathcal{S}^\pm$, whose construction has been explicitly discussed in subsections \ref{sec2.3} with appendix \ref{appA} and \ref{sec2.6}, respectively. We also constructed the winding-one representatives of the single-trace $T\bar{T}$ and $J\bar{T}$ operators in the decoupled theory. 

In section \ref{sec3}, we extended the discussion to single-trace $T\bar{T}$ deformed string theory on $AdS_3\times \mathcal{N}$ at $k=1$. We argued that in the long-strings decoupling limit of single-trace $T\bar{T}$-deformed string theory on $AdS_3\times \mathcal{N}$ at $k=1$, the spacetime theory is the symmetric product of $T\bar{T}$ deformed $\mathbb{R}\times \mathcal{N}$. As a couple of examples, we computed the two-point correlation functions in momentum space of spin-1 conserved currents, and of the spin-2 component of the stress-energy tensor, in the spacetime theory. In the absence of an independent field theory computation, it remains a non-trivial prediction from string theory.~\footnote{Note however that the large-momentum behavior of the correlation functions \eqref{kk} and \eqref{tt}, which is proportional (after stripping off polynomial factors in $p,\bar p$ and introducing a renormalization scale, $\mu$) to
\begin{equation}
    \frac{1}{\sin\left(\pi\frac{\alpha'}{2}\lambda p\bar{p}\right)}\left({p\bar{p}\over\mu^2}\right)^{-\frac{\alpha'}{2}\lambda p\bar{p}}~,
\end{equation} 
is the same as the large-momentum behavior of two-point functions of scalar primary operators found in~\cite{Aharony:2023dod} using the JT-gravity formulation of $T\bar{T}$-deformed CFT, as well as in string theory,~\cite{Cui:2023jrb,Giveon:2023gzh}. It should also be noted that according to~\cite{Aharony:2023dod}, as in~\cite{Asrat:2017tzd}, we ignored problems with regularization and renormalization in Fourier transforms such as~\eqref{fta}, when $h\to h_{p^2}$. A complete understanding of these issues is beyond the scope of this note.}

Although the emphasize in this paper has been on string theory on $AdS_3\times\mathcal{N}$ that gives rise to $N=(2,2)$ SCFT in spacetime, the analysis in this note can be readily generalized to any fermionic string on (deformed) $AdS_3\times{\cal N}$ with ${\cal N}$ an $N=1$ SCFT with $c_{\cal N}=9/2$, \eqref{k}.
Various universal properties will remain, \eg the existence of holomorphic $\partial_x\phi$ in spacetime, in the $k=1$ limit, and its on-shell worldsheet construction.
But other properties require further study. 
For instance, in the $k=1$ limit, the spacetime theory includes also a free fermion, $\psi_\phi$, which is related to $\phi$ by an emergent supersymmetry in the $\mathbb{R}_\phi$ of the long-string theory,~\cite{Chakraborty:2025nlb}.
In the $N=(2,2)$ supersymmetric case, the on-shell worldsheet operator corresponding to $\psi_\phi$ is the ${\cal S}^++{\cal S}^-$ combination of the ${\cal S}^\pm$ in \eqref{ss}.
Needless to say that it would be interesting to construct the on-shell worldsheet operator corresponding to $\psi_\phi$ in the generic case. 
This, as well as other generic properties of the fermionic strings on  (deformed) $AdS_3\times\mathcal{N}$ with $k=1$,
are left for future work.

The single-trace $T\bar{T}$  holography at $k=1$ discussed in this paper is focused on the particular case where the geometry is asymptotically a flat spacetime with a linear dilaton in the UV, smoothly capped off by a Poincar\'e patch of global $AdS_3$ or massless BTZ in the IR. Yet, the discussion in this paper can be generalized to the case where the IR geometry is global $AdS_3$, instead, following \eg the techniques discussed in~\cite{Chakraborty:2024mls}. In the long-strings decoupling limit, the dual spacetime theory is the  SCFT on $(\mathbb{R}\times \mathcal{N})^p/S_p$ in its NS sector. Finite temperature generalizations (a.k.a. massive BTZ caps in the IR) are also straightforward, \eg along the lines in \cite{Chakraborty:2024ugc,Giveon:2024sgz}.

Finally, there are non-Lorentz invariant generalizations of single-trace $T\bar{T}$ holography: deformations of string theory on $AdS_3\times \mathcal{N}$ (for generic $k$) by a general linear combination of single-trace $T\bar{T}, J\bar{T}$ and  $T\bar{J}$ \cite{Chakraborty:2019mdf,Apolo:2021wcn}. The deformed worldsheet theory is well understood, and the spectrum of long-strings can be shown to be identical to the spectrum of the symmetric product of  $T\bar{T}+ J\bar{T}+T\bar{J}$ deformed long-string CFT$_2$,~\eqref{sw}. The $k=1$ decoupling of the long-strings is straightforwardly applicable here as well. The resulting spacetime theory would be a $p$-fold symmetric product of  $T\bar{T}+ J\bar{T}+T\bar{J}$ deformed $\mathbb{R}\times \mathcal{N}$.
It would thus be interesting to obtain from string theory on deformed $AdS_3\times{\cal N}$ explicit predictions for $T\bar{T}+ J\bar{T}+T\bar{J}$ deformed CFT$_2$.

%\vspace{10mm}

\section*{Acknowledgements} 

We thank D. Kutasov for collaboration on many parts of this paper.
The work of SC received funding from the Department of Physics at The Ohio State University. SC would like to thank the University of Chicago for its hospitality during part of this work.
The work of AG was supported in part by the ISF (grant number 256/22).

\vspace{10mm}
\appendix

\section{Extending results of \cite{Balthazar:2021xeh} on $\partial_x\phi$ and $\partial_{\bar{x}}\partial_x\phi$ to $k=1$}\label{appA}

Repeating the manipulations in sections 4.1 and 7.2 of \cite{Balthazar:2021xeh} for $k=1$, one finds the following.
The operator $\partial_x\phi$ in the spacetime theory corresponds on the worldsheet of the superstring on $AdS_3\times \mathbb{T}^3$
(and/or of any fermionic string on $AdS_3$ with $k=1$)
to the on-shell operator
\be\label{aa}
\partial_x\phi\,\longleftrightarrow\, e^{-\varphi-\bar\varphi}e^{i(H_{sl}+\bar H_{sl})}(\partial\varphi+i\partial H_{sl})\Phi^{(-1)}_{{1\over 2};-{1\over 2},{1\over 2}}~.
\ee
Namely, as in (4.27) of \cite{Balthazar:2021xeh},
we find that the worldsheet theory contains an operator with spacetime scaling dimension $(1,0)$ (the r.h.s. of \eqref{aa})
that corresponds in the long-string theory (the block of the symmetric product CFT) to the operator $\partial_x\phi$ (the l.h.s.).
Now, a similar calculation to that in (7.4),(7.5) of \cite{Balthazar:2021xeh}, reveals that
the operator $\partial_{\bar{x}}\partial_x\phi$ in $(\mathbb{R}\times \mathbb{T}^3)^p/S_p$ corresponds on the worldsheet to the string state obtained in the limit 
\be\label{ab}
\partial_{\bar{x}}\partial_x\phi\,\longleftrightarrow\, \lim_{s\to 0}ise^{-\varphi-\bar\varphi}e^{i(H_{sl}+\bar H_{sl})}(\partial\varphi+i\partial H_{sl})(\bar\partial\varphi+i\bar\partial H_{sl})\Phi^{(-1)}_{{1\over 2}+is;-{1\over 2}-s^2,-{1\over 2}-s^2}~,
\ee
which for real radial momentum in $\phi$, $s\in \mathbb{R}$, as required to obtain the string theory dual of the free symmetric product, gives
\be\label{ac}
\partial_{\bar{x}}\partial_x\phi=0~.
\ee

\noindent
A few comments are in order:
\begin{itemize}
\item
While in the $\epsilon\to 0$ limit of (7.4) in \cite{Balthazar:2021xeh} one approaches an LSZ pole $\sim 1/\epsilon$, that cancels the overall $\epsilon$ factor in the limit, giving rise to the {\it normalizable} operator in (7.5),
$e^{-\varphi-\bar\varphi}e^{i(H_{sl}+\bar H_{sl})}(\partial\varphi+i\partial H_{sl})(\bar\partial\varphi+i\bar\partial H_{sl})\Phi^{(-1)}_{{1\over 2};-{1\over 2},-{1\over 2}}$, which implies that $\partial_{\bar{x}}\partial_x\phi\neq 0$, here, since $\epsilon=is$ is imaginary, one doesn't approach a pole in the limit, thus getting zero in \eqref{ab}, instead, and consequently $\partial_x\phi$ is holomorphic,~\eqref{ac}.
\item
To allow $j={1\over 2}+is$ with real $s$ in (7.3) of \cite{Balthazar:2021xeh}, it is important that $m$ and $\bar m$ will remain real on-shell also for $s\neq 0$; while for $k\neq 1$ it ain't so (see (7.2)), remarkably, for $k=1$, $m$ and $\bar m$ are real for any $s$: from (2.60) in~\cite{Balthazar:2021xeh}, one finds $m=\bar m=-{1\over 2}-s^2$ on shell (as indicated in \eqref{ab}).
\item
Various choices of normalization factors in \cite{Balthazar:2021xeh} that involved $1-k$ (\eg in (7.3)) were harmless there (since $1-k>0$ in~\cite{Balthazar:2021xeh}), but here $1-k=0$, and one should avoid such $1-k$ factors in the choice of normalization (as is done in~\eqref{aa}).
\item
An equivalent way to obtain \eqref{ac} is directly from the extension of (7.1) in \cite{Balthazar:2021xeh} to the $k=1$ case,
\be\label{ad}
\partial_{\bar{x}}\partial_x \phi\,\longleftrightarrow\,\left\{\bar Q_{BRST},e^{-\varphi-2\bar\varphi}(\partial\varphi+i\partial H_{sl})\bar\partial\bar\xi\Phi^{(-1)}_{{1\over 2};-{1\over 2},{1\over 2}}\right\}~,
\ee
in the following way.
First, we regard $\Phi^{(-1)}_{{1\over 2};-{1\over 2},{1\over 2}}$ as being obtained in the limit of delta-function normalizable operators, when approaching the boundary of the continuum, namely, by taking the limit where the real $\phi$ momentum goes to zero:
$\Phi^{(-1)}_{{1\over 2};-{1\over 2},{1\over 2}}\equiv\lim_{s\to 0}\Phi^{(-1)}_{{1\over 2}+is;-{1\over 2}-s^2,-{1\over 2}-s^2}$.
This way of thinking about $\Phi^{(-1)}_{{1\over 2};-{1\over 2},{1\over 2}}$ in \eqref{ad} is required for this operator to live in the infinite $\phi$ regime.
Now, the subtlety discussed following (7.1) in~\cite{Balthazar:2021xeh} does not apply, namely, the operator on the r.h.s. of \eqref{ad} is BRST exact,
and thus $\partial_{\bar{x}}\partial_x\phi=0$ in the BRST cohomology.
\end{itemize}

\section{FZZ duality}\label{appB}

The FZZ duality in the $(m,\bar m)$ basis reads (see \eg (2.62) in \cite{Balthazar:2021xeh} and refs. therein)
\be\label{fzz}
\Phi^{(w)}_{j;-j,-j}(z,\bar z)\simeq\Phi^{(w-1)}_{{k+2\over 2}-j;{k+2\over 2}-j,{k+2\over 2}-j}(z,\bar z)~,
\ee
so, formally, in the $(x,\bar x)$ basis, \cite{Maldacena:2001km} (see also section 4.2 in \cite{Iguri:2022eat}~\footnote{In \cite{Iguri:2022eat}, a suggestion for a {\it local} definition of $\Phi^{w,j}_h(x,\bar x;z,\bar z)$ is provided for any $w\geq 1$, generalizing the one proposed in~\cite{Maldacena:2001km}.} and refs. therein),
\be\label{fzzx}
\Phi^{w,j}_{h=-j+{k+2\over 2}w}(x,\bar x;z,\bar z)\simeq\Phi^{w-1,{k+2\over 2}-j}_{h={k+2\over 2}-j+{k+2\over 2}(w-1)}(x,\bar x;z,\bar z)~,\qquad w\geq 1~.
\ee
The ${\cal K}^a(x)$ in \eqref{ks} and \eqref{kwone} are thus a particular dual pair (which amounts to the case $w=1,j=1/2,h=1$and $k=1$ in \eqref{fzzx}).~\footnote{While it ain't clear if $\Phi^{w,j}_h(x,\bar x;z,\bar z)$ are generically good local operators, when $w\geq 1$, the way that the particular $\Phi^{w=1,j=1/2}_{h=1}(x,\bar x;z,\bar z)$, in the $k=1$ case, is obtained in the limit~\eqref{lims}, gives rise to a  well defined local operator.}

\section{Infinite phase shift at $k\to 1$}\label{appC}

In the decoupled long-strings theory at $k=1$, in particular, we regarded (real) positive and negative radial momenta as independent.
To illustrate what’s special about $k=1$, which allows that, consider the two-point function defined
in eq. (4.1) of \cite{Giveon:2001up}. The principal continuous series corresponds to $2h-1=is$, where $s\in\mathbb{R}$ is proportional to the radial momentum. The two-point function $D(h)$ is given by equations (4.36), (4.37) in that paper, where $t=2-k=-1$ for the case of interest. For that value of $t$, the denominator in (4.37) is singular.

There are two possible ways to deal with this singularity. One is to keep the constant $\lambda$ finite in the limit $t\to -1$. The prefactor $\nu^{2h-1}$ becomes in this case an infinite phase shift, and thus the scattering from the interior of $AdS_3$ that it represents can be ignored. The second way is to rescale $\lambda$  such that the factor $\nu$ (4.37) remains finite in this limit; this gives a finite scattering amplitude. 
The former limit is in harmony with the positive and negative (real) radial momenta being independent in the decoupled $k=1$ theory at infinite $\phi$.

\vspace{10mm}

%\bibliography{ref}\bibliographystyle{JHEP}

\providecommand{\href}[2]{#2}\begingroup\raggedright\endgroup

\end{document}